# Breaking the bonds of weak coupling: the dynamic causal modelling of oscillator amplitudes

Erik D. Fagerholm[1,*], Rosalyn J. Moran[1], Inês R. Violante[2], Robert Leech[1,§], Karl J. Friston[3,§]

[1] Centre for Neuroimaging Sciences, Department of Neuroimaging, IOPPN, King's College London
[2] School of Psychology, Faculty of Health and Medical Sciences, University of Surrey
[3] Wellcome Trust Centre for Neuroimaging, Institute of Neurology, University College London

§ These authors contributed equally to this work

* Corresponding author: erik.fagerholm@kcl.ac.uk

**Abstract**

Models of coupled oscillators are useful in describing a wide variety of phenomena in physics, biology and economics. These models typically rest on the premise that the oscillators are weakly coupled, meaning that amplitudes can be assumed to be constant and dynamics can therefore be described purely in terms of phase differences. Whilst mathematically convenient, the restrictive nature of the weak coupling assumption can limit the explanatory power of these *phase-coupled* oscillator models. We therefore propose an extension to the weakly-coupled oscillator model that incorporates *both* amplitude and phase as dependent variables. We use the bilinear neuronal state equations of dynamic causal modelling as a foundation in deriving coupled differential equations that describe the activity of both weakly and strongly coupled oscillators. We show that weakly-coupled oscillator models are inadequate in describing the processes underlying the temporally variable signals observed in a variety of systems. We demonstrate that *phase-coupled* models perform well on simulations of weakly coupled systems but fail when connectivity is no longer weak. On the other hand, using Bayesian model selection, we show that our *phase-amplitude coupling* model can describe non-weakly coupled systems more effectively despite the added complexity associated with using amplitude as an extra dependent variable. We demonstrate the advantage of our phase-amplitude model in the context of model-generated data, as well as of a simulation of inter-connected pendula, neural local field potential recordings in rodents under anaesthesia and international economic gross domestic product data.

**Introduction**

Oscillations can be observed across a variety of natural systems. However, *pure* oscillations that exhibit a stable limit cycle – a closed trajectory in phase space towards which nearby curves spiral – serve only as a mathematical construct and are not observed in real physical or biological systems. In the brain for example, quasi-stationary and quasi-band-limited processes are the norm (1). Neural oscillations may facilitate information exchange across the brain (2-4), creating a dimension of signalling at a more coarse-grained network level (i.e., a mean field description) (5-7), compared with spiking activity at the level of individual neurons. At present the most compelling evidence for the role of oscillations in human brain function appeals to effects on movement and memory when phase resetting methods are used to disturb ongoing oscillatory activity; for example, in Parkinson's disease (8) and sleep studies (9, 10). While the specific mechanisms underlying oscillations may be complex, recovering interactions among individual oscillators remains a crucial aspect in establishing their role in neuronal, cognitive and pathogenic settings (4).

Often interactions amongst oscillators can be approximated by applying the weak coupling assumption. This restricts motion in phase space, such that deviations from the limit cycle are considered to be negligible in the equations of motion. The small off-limit cycle deviation can be visualized as a narrow torus in phase space that defines the boundaries of allowable states (Fig. 1A).

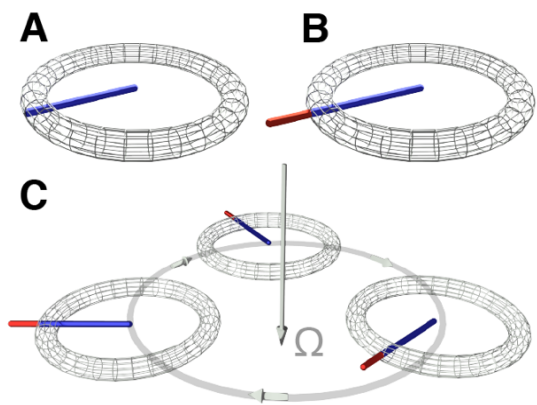

*Fig. 1: Oscillator dynamics in phase space,* **(A)** *A torus surrounding an oscillator's limit cycle (radius shown by the blue line) in phase space,* **(B)** *Same as (A), except dynamics are now allowed to evolve beyond the torus (shown by the red line),* **(C)** *Three coupled oscillators, with a global angular frequency shown by the vector $\Omega$.*

Under the weak coupling assumption, oscillators evolve with constant amplitudes, and are therefore only able to alter each other's phases (11). Upon first inspection, omitting amplitude dynamics seems like an attractive option when simulating complex dynamical systems – as one reduces the description to phase as the only dependent variable. Models based on the weak coupling assumption have become by far the most commonly used in neuroimaging studies, where dimensionality reduction is one way of rendering analyses more computationally tractable. Owing to this increased tractability, phase-only models have allowed for large-scale modelling of dynamic repertoires such as multistability, transience and criticality, with spatial embedding across the cortical sheet (12, 13). Phase-only models have also been shown to be able to capture patterns of macroscopic neural dynamics consistent with those reported with magnetoencephalography (MEG) and functional magnetic resonance imaging (fMRI) (14, 15), showing similar performance to more involved approaches (16).

One of the most widely used phase-only models is the Kuramoto model (17), which was originally developed by studying chemical phenomena, but has also been used to explore oscillations in physical, biological and social systems (18, 19). The model can capture a range of relatively complicated dynamics, whilst being sufficiently simple to allow for the modelling of a large number of interacting components. The Kuramoto model assumes that different interacting components of a system can be approximated as weakly-coupled oscillators, operating on limit cycles with specific natural frequencies. However, the price which phase-only descriptions – such as the Kuramoto model – pay for treating the limit cycle (and its toroidal shell) as a fixed boundary, is that they leave the potentially interesting behaviour beyond the limit cycle unexplored and uncharacterised. For this reason, we set out to derive a model in which we allow system states to evolve beyond the limit cycle (Fig. 1B). We obtain a full description of the dynamics by measuring the amplitudes and relative phase differences of the individual oscillators, as well as the angular frequency of the system as a whole (Fig. 1C). We then use Bayesian methods to determine whether this ability to accommodate phase-amplitude dynamics is advantageous for modelling purposes, given the increased complexity compared with phase-only descriptions.

We begin with the bilinear neuronal state equations underlying dynamic causal modelling (DCM) for fMRI (20). We make the minimum adaptations necessary to these equations to achieve the broadest possible description of dynamics both on and beyond the limit cycle. We show that this generalisation reduces to the Kuramoto model in the limit of the weak coupling assumption. We further show that the generalised DCM equation allows an uncoupled oscillator to retain intrinsic periodicity in the absence of external input, thereby accounting for a more realistic description of natural oscillatory systems. We test the generalised DCM equation with data that is a) model-generated, b) derived from a simulation of coupled pendula, c) taken from local field potential (LFP) recordings in rodents at different levels of anaesthesia, and d) collected from gross domestic product (GDP) records across different countries. In all four cases, we use Bayesian model reduction to test for evidence for the relative importance of phase and amplitude coupling effects. We show that taking both phase and amplitude into account is advantageous in modelling natural phenomena.

**Results**

**Model-generated data**: here, we offer proof of principle that a coupled oscillator model with both phase and amplitude states can be inverted using generalised filtering (Matlab code: https://github.com/Phase-Amplitude-Coupling/Code.git). Generalised filtering refers to a variational Bayesian scheme that estimates parameters and hyperparameters as well as states in several degrees of motion. It is a fairly standard scheme used routinely in dynamic causal modelling and can be regarded as a generalisation of Bayesian filtering under the Laplace assumption. To establish construct validity, we generated data from a system of three interacting regions that are coupled in terms of their phase and amplitude in a hierarchical fashion; i.e. one region connects to a subordinate oscillator, which in turn connects to another subordinate oscillator (Fig. 2A). The ensuing timeseries were then subject to generalised filtering under competing models (i.e., phase coupling versus phase-amplitude coupling) to produce estimates (i.e., posterior densities) of the coupling parameters and the evidence for each model. Because the parameters and models generating the data are known, one can use the posterior estimates to establish the estimability of parameters and identifiability of competing models.

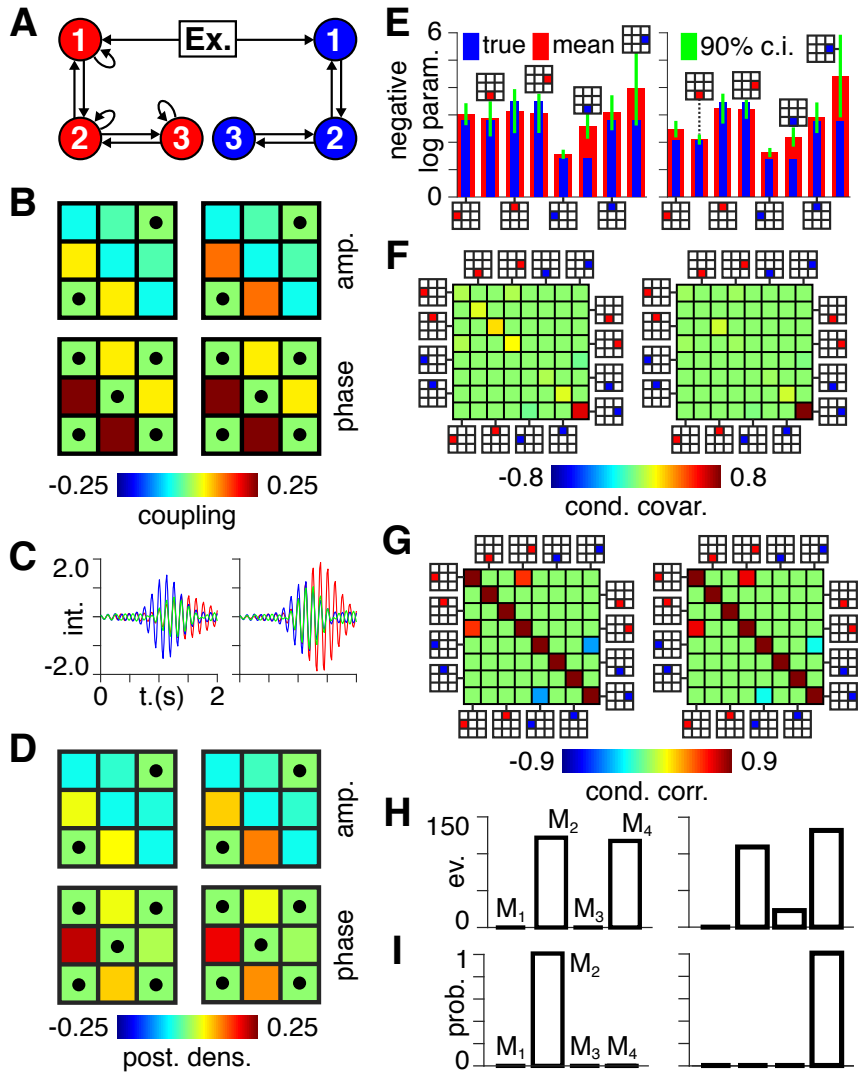

**Fig. 2: synthetic data. (A)** an exogenous input (Ex.) connects to node 1 of both the amplitude (red) and phase (blue) networks. **(B)** Coupling strengths between nodes of the amplitude (top row) and phase (bottom row) networks for a phase-only (left column) and phase-amplitude (right column) coupling scenario. Black dots indicate non-existent coupling. **(C)** Model-generated signals for phase-only (left) and phase-amplitude (right) coupling, where the driven node is shown in blue. **(D)** A posteriori estimates of the amplitude (top row) and phase (bottom row) matrices for phase-only (left column) and phase-amplitude (right column) coupling scenarios. Black dots indicate non-existent coupling. **(E)** All values shown in negative logarithmic space. Mean estimated parameter values (red) vs. true parameter values (blue), together with 90% confidence intervals (green) for phase-only (top) and phase-amplitude (bottom) coupling. The first and last four parameters are the amplitude (red) and phase (blue) coupling elements indicated by the insets, respectively. **(F)** Conditional co-variance values for phase-only (left) and phase-amplitude (right) coupling. The first and last four parameters are the amplitude (red)

*and phase (blue) coupling elements indicated by the insets, respectively.* **(G)** *Conditional correlations for phase-only (left) and phase-amplitude (right) coupling. The first and last four parameters are the amplitude (red) and phase (blue) coupling elements indicated by the insets, respectively.* **(H)** *Approximate log model evidence for phase-only (left) and phase-amplitude (right) coupling following Bayesian model reduction for models $M_1$: no phase or amplitude coupling, $M_2$: phase coupling only, $M_3$: amplitude coupling only, and $M_4$: both phase and amplitude coupling.* **(I)** *Probabilities derived from the log evidence for the models in (E).*

In detail, we simulated data using equations [7] and [8] in Methods for both phase-only and phase-amplitude coupling (Fig. 2B). The resultant time series (Fig. 2C) were Hilbert transformed into their analytic signals, which subsequently constitute the data feature used for model inversion (data fitting), using the generalised filtering scheme. This procedure optimises the states, parameters and hyperparameters of the model for any given multivariate time series. By assuming fairly precise priors on the amplitude of random fluctuations one can recover the parameters (Fig. 2D,E) – their maximum a posteriori (MAP) estimates and their posterior covariation (Fig. 2F) and correlation (Fig. 2G) – and use the approximate marginal log probability (free energy) of alternative models (Fig. 2H) for subsequent Bayesian model comparison. In particular, we used Bayesian model reduction to assess the probability for models with 1) no phase or amplitude coupling, 2) phase coupling only, 3) amplitude coupling only, and 4) both phase and amplitude coupling (Fig 2I).

We compared the estimated parameter values (Fig. 2D,E) to the values used to generate the data (Fig. 2B,E). Prior to model reduction we looked for an effect of phase-amplitude coupling in the estimated coupling matrices and found that non-zero amplitude coupling effects are captured a posteriori (Fig. 2D,E). Importantly, the inversion recovers stronger amplitude matrix elements for the phase-amplitude coupling data compared with phase-only coupling. For the phase estimates, posterior differences were weaker (Fig. 2D,E), in line with the generative models used (Fig. 2B). Figure 2E shows posterior parameter estimates, true parameter values and 90% Bayesian credible intervals for the optimal (i.e., best) models. The first two parameters in Figure 2E are the amplitude coupling matrix elements that were varied in order to switch between phase-only and phase-amplitude coupling scenarios (Fig. 2I). The first of these parameters does not fall within the 90% interval for the phase-amplitude coupling scenario. However, the estimated parameter value remains conservative, in that it lies between the true and the prior value (-3 in logarithmic space). This reflects a well-known effect of shrinkage priors in variational Bayes (21). We then evaluated conditional co-variances (Fig. 2F) and conditional correlations (Fig. 2G) between parameters. One can see that the most correlated parameters (indicative of lesser identifiability) are among the amplitude rather than phase coupling matrix elements. Bayesian model reduction correctly identified the structure of the generative model – selecting the correct model (out of four) for data generated with and without amplitude coupling (Fig. 2H,I). This means that, under conditions of actual weak (phase-only) coupling, the Bayesian model reduction (which incorporates complexity penalization) correctly identifies the simpler model.

**Coupled pendula simulation**: here, we used the same procedure as for the model-generated data (Matlab code: https://github.com/Phase-Amplitude-Coupling/Code.git) to investigate phase-amplitude dynamics in a simulation of three coupled pendula (Fig. 3A).

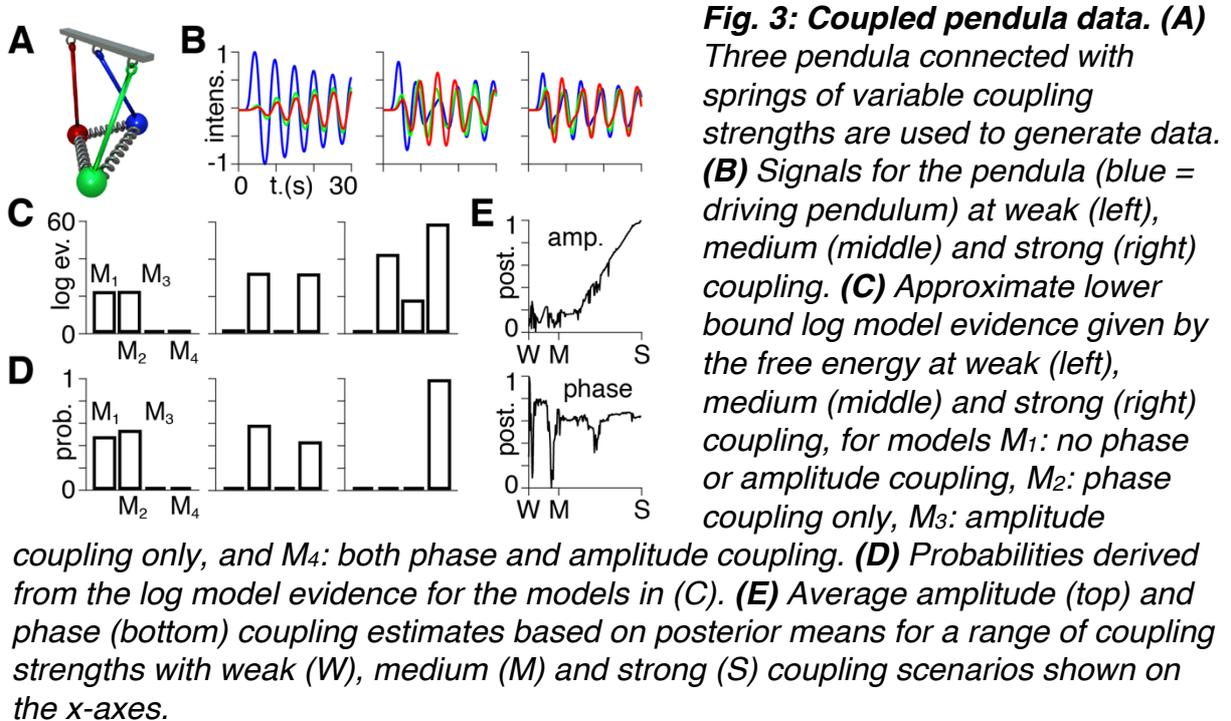

*Fig. 3: Coupled pendula data. (A) Three pendula connected with springs of variable coupling strengths are used to generate data. (B) Signals for the pendula (blue = driving pendulum) at weak (left), medium (middle) and strong (right) coupling. (C) Approximate lower bound log model evidence given by the free energy at weak (left), medium (middle) and strong (right) coupling, for models $M_1$: no phase or amplitude coupling, $M_2$: phase coupling only, $M_3$: amplitude coupling only, and $M_4$: both phase and amplitude coupling. (D) Probabilities derived from the log model evidence for the models in (C). (E) Average amplitude (top) and phase (bottom) coupling estimates based on posterior means for a range of coupling strengths with weak (W), medium (M) and strong (S) coupling scenarios shown on the x-axes.*

In this setup, one pendulum is driven by an exogenous Gaussian input lasting for 5 seconds and the resultant motion of all three pendula are recorded for a total of 30 seconds at a sampling rate of 60 Hz. The coupling strengths between the pendula were then progressively increased by varying the spring constants in the simulation. The ensuing coupling strengths ranged between the weak, medium and strong, illustrated in terms of the system's response (Fig. 3B). These signal outputs were then Hilbert transformed to obtain phase and amplitude information from the analytic signals for generalised filtering.

Bayesian model reduction was performed following model inversion to calculate posterior parameter estimates and model evidence (Fig. 3C) as above using the same four models: neither phase nor amplitude coupling, phase-coupling only, amplitude-coupling only, or both phase and amplitude coupling. We found that the weak coupling data were best explained by the 'no phase or amplitude coupling' and 'phase coupling only' models. The medium coupling data were best explained by the 'phase coupling only' and 'both phase and amplitude coupling' models. The strong coupling data were best explained by the 'both phase and amplitude coupling' model.

The average posterior connectivity estimates of amplitude coupling (pooled over elements of the coupling matrix) increases with the actual coupling strength (r=0.93, p<0.001, Spearman) (Fig. 3E, top). These averages have been normalised between zero and unity in the figure, but individual estimates were occasionally negative. No relationship was observed between the average phase coupling matrix elements and spring coupling strength (Fig. 3E, bottom).

**Neuroimaging data**: here, we used the same method to assess the importance of phase and amplitude coupling under different levels of Isoflurane anaesthesia in two-channel LFP recordings from primary and secondary rodent auditory cortices (22).

The LFP data (Fig. 4A) were split into ten non-overlapping time windows of equal length, each of which was Hilbert transformed to obtain the phase and amplitude from the analytic signal. The model used to explain these data is a simplified version of the model used to simulate timeseries above (Fig. 2A) – with two bi-directionally connected nodes, both of which connect to a (white noise) exogenous input, modelling afferent neuronal fluctuations.

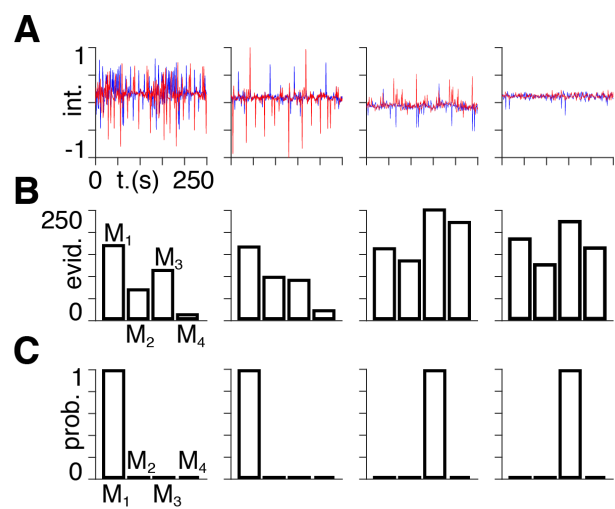

*Fig. 4: LFP data in rodent auditory cortex. Four doses of isoflurane were applied intraperitoneally. The difference in dosage is associated with altered local field potentials in A1 (blue) and the posterior auditory field (red) shown left to right for 1.4, 1.8, 2.4 and 2.8% isoflurane **(A)** Two-channel normalized signal intensity. **(B)** Approximate log model evidence given by the free energy for models $M_1$: no phase or amplitude coupling, $M_2$: phase coupling only, $M_3$: amplitude coupling only, and $M_4$: both phase and amplitude coupling. **(C)** Probabilities derived from the log model evidence in (B).*

Bayesian model reduction was performed following inversion to evaluate the evidence for the four models of coupling (Fig. 4B). The resulting log evidence values were summed across consecutive time windows to calculate the associated probabilities, under the assumption that the consecutive windows constitute conditionally independent data (Fig. 4C). The results of this model suggest that the two lowest doses of anaesthesia are best explained by the 'no phase or amplitude coupling' model. Crucially, the data from the two highest doses of anaesthesia are best explained by the 'only amplitude coupling' model. This is an interesting result that reflects the importance of amplitude coupling, in relation to phase only coupling.

**Economic data**: here, we applied the same procedures to explore the change in coupling dynamics between the economies of different countries before and after WWII, in terms of their GDP per capita. We used a modified version of our model (Fig. 2A) with five fully connected nodes, all of which received exogenous input, to account for random fluctuations in generating the timeseries.

We used the GDP per capita records for the United States, Germany, the United Kingdom, Japan and France (23). We divided the detrended data into two equal periods of time preceding and following WWII, one between 1870 and 1938 and the other between 1946 and 2014 (Fig. 5A).

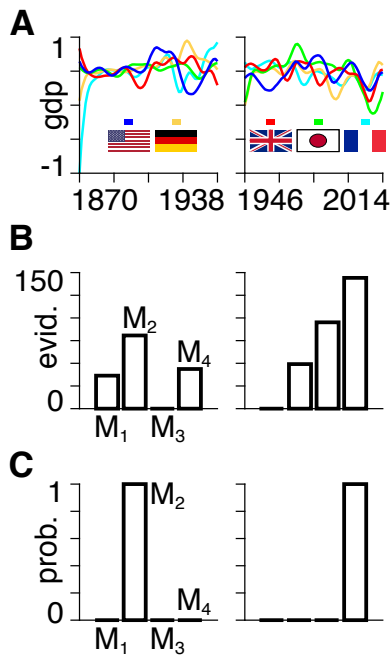

*Fig. 5: GDP per capita. Detrended, smoothed and normalized GDP per capita records divided into the 68 years preceding (left) and following (right) WWII. (A) The data for the different countries are represented by colours corresponding to the flags in the insets. (B) Approximate log model evidence given by the free energy for models $M_1$: no phase or amplitude coupling, $M_2$: phase coupling only, $M_3$: amplitude coupling only, and $M_4$: both phase and amplitude coupling. (C) Probabilities based on the log model evidence for the models in (B).*

The data were Hilbert transformed to obtain the phase and amplitude. As above, we applied generalised filtering and Bayesian model reduction to estimate coupling parameters and log model evidence (Fig. 5B) to evaluate the associated probabilities (Fig. 5C). The pre and post-WWII eras are described most accurately by the 'phase coupling only' and 'both phase and amplitude coupling' models, respectively.

**Discussion**

This paper introduces a coupled oscillator model that does not rely on the weak coupling assumption and can therefore model both phase and amplitude coupling. We incorporated this model into the DCM framework and established that it can be used to make inferences about both directed phase and amplitude coupling over a range of various data sources. This provided proof of principle that one can model a broader dynamic repertoire by allowing oscillator amplitudes to evolve beyond their limit cycles, instead of constraining the amplitude to remain constant, as per the weak coupling assumption. We also demonstrated that our phase-amplitude coupling model outperforms phase-only models, despite the increased model complexity associated with the inclusion of amplitude as an extra dependent variable. In addition to establishing face validity using data that were generated analytically, we also used data from a physics engine for coupled pendula with known coupling strengths. Finally, we used the model of phase-amplitude coupling to ask whether amplitude coupling, above and beyond phase coupling, is a feature of empirical data. We confirmed this to be the case using neural data acquired at different levels of anaesthesia, and economic data from different countries before and after WWII. Although the particular model of phase and amplitude coupling used in this paper is sufficient for our purposes, the coupled differential equations that constitute this model, should be thought of as one possible (i.e., minimal) model of phase and amplitude in coupled oscillator models. For example, there is evidence that sinusoids may not best characterise periodic neural signals (1). Alternative formulations of the model used here could therefore explore different waveforms.

In the neural data, we observed a shift from there being no coupling between the two auditory regions, to amplitude playing a prominent role – as the level of anaesthesia increases. This result is consistent with studies showing that anaesthesia leads to neuronal dynamics that are less complex (24) and that operate further from a phase transition (25). Contrasting our present analysis with previous approaches, applying DCM to this dataset illustrates the strengths and weaknesses of modelling neural systems as coupled oscillators. The current approach allows us to explicitly quantify the separate contributions from phase and amplitude – and does so in a manner which is computationally tractable and so can scale across many regions. On the other hand, other approaches applied to the same dataset (22) employ more biologically grounded biophysical (neural mass) models, which support inferences about the underlying biological parameters (e.g., GABAergic and glutamatergic mechanisms). The appropriateness of using the proposed phase and amplitude coupled oscillator model will clearly depend on the specific question and the mechanistic scale under consideration.

Our GDP analysis demonstrates how one can analyse a completely different type of data – compared with neuronal timeseries – using the same phase-amplitude coupled oscillator model. Economies, measured as deviations in gross domestic product from their long-term trends, have long been recognised to follow cycles, first studied by Sismondi in the early 19th Century. However, while phase-only models have been applied to economic data (26), economic cycles do not have consistent or stable periodicity. Incorporating amplitude, as well as phase, allows us to model measured time series comprising non-uniform oscillations and to make inferences about the relative role of phase and amplitude over different time periods. For the five countries considered we observed that, prior to WWII, the best model was the one based purely on phase-coupling between countries, suggesting a more periodic oscillatory scenario. On the other hand, post WWII, the model with the highest evidence included both phase and amplitude, suggesting more complex coupling among countries in the modern age. Our approach has the potential to quantify intrinsic and exogenous contribution to economic cycles, an issue which has long been discussed (27).

Models that attempt to explain the mechanisms underlying neuronal coupling proffer hypotheses of how these features emerge and might break down in neuropsychiatric disorders. Using a model such as ours allows for formal characterisation of how different neuronal dynamics, cognitive tasks or sources of individual variability (e.g., related to pathology) are driven by alterations to amplitude or phase, or a combination of both. In addition, by modelling neuronal data in terms of both phase and amplitude, we can potentially model aperiodic patterns, which are thought to play a prominent role in functional integration in the brain (4, 28). Synchronisation is a phenomenon thought to be associated with cognitive functions, including attention and memory, and one which can reach pathophysiological levels in epilepsy and movement disorders (29-31). Neural oscillations therefore provide an example of how assessing amplitude and phase contributions could be beneficial.

**Methods**

**Equations of motion:** here we derive differential equations describing a driven system of weakly or strongly coupled oscillators. Considering the bilinear neuronal state equations underlying DCM for fMRI (20) . These equations were originally chosen as the simplest second-order (Taylor approximation) to any ordinary differential equations state space model:

$$\frac{dz}{dt} = \left(A + \sum_j u_j B^j\right)z + Cu \qquad [1]$$

We set out to make the minimum changes to [1] that allow for the broadest possible description of coupled oscillator dynamics occurring both on and beyond their limit cycles. As a first step, we re-define the dependent variable $z$ in [1] to be a complex variable $z_c$, such that:

$$z(t) \;\rightarrow\; z_c\big(r(t), \theta(t)\big) = r(t)e^{i\theta(t)} \qquad [2]$$

Where $r$ is the amplitude and $\theta$ is the phase. $r$ comprises both a weak-coupling amplitude $r_W$ (i.e. the limit cycle radius, Fig. 1A), and a strong-coupling amplitude $r_S$ (Fig. 1B). $r_S$ is defined as the perpendicular distance extending beyond the limit cycle radius (32), such that:

$$r(t) = r_S(t) + r_W \qquad [3]$$

We re-write [1] using [2], and make the two additional modifications shown in red:

$$\frac{dz_c}{dt} = \left(\boldsymbol{X}A + \sum_j u_j B^j + \boldsymbol{Y}\right)z_c + Cu \qquad [4]$$

Where $\boldsymbol{X} = \frac{r - r_W}{r}$ and $\boldsymbol{Y} = i\left(\Omega + \sum_j A^j sin(\theta_j - \theta)\right)$

Equating the real and imaginary components of [4], we obtain the following coupled differential equations:

$$\frac{d\theta}{dt} = \Omega + \sum_j A^j \sin(\theta_j - \theta) - \frac{Cu\sin\theta}{r} \quad [5]$$

$$\frac{dr}{dt} = A(r - r_W) + r\sum_j u_j B^j + Cu\cos\theta \quad [6]$$

We require two separate intrinsic coupling matrices for phase and amplitude that can be manipulated independently for the purpose of Bayesian model comparison. Furthermore, we would like the phase and amplitude inter-dependence in [5] and [6] to be built into the two intrinsic coupling matrices. We therefore replace the $A$ matrix in [5] by an amplitude-dependent matrix $A_r$, in which matrix element $a_{ij}$ is multiplied by $e^{-|r_j - r_i|}$. Similarly, we replace the $A$ matrix in [6] by a phase-dependent matrix $A_\theta$, in which matrix element $a_{ij}$ is multiplied by $\cos(\theta_j - \theta_i)$. The forms of the element-wise factors in $A_r$ and $A_\theta$ ensure that the unmodified coupling strength $a_{ij}$ is recovered when $r_j = r_i$ and $\theta_j = \theta_i$, respectively. We therefore write the final forms of the equations of motion as follows:

$$\frac{d\theta}{dt} = \Omega + \sum_j A_r^j \sin(\theta_j - \theta) - \frac{Cu\sin\theta}{r} \quad [7]$$

$$\frac{dr}{dt} = A_\theta(r - r_W) + r\sum_j u_j B^j + Cu\cos\theta \quad [8]$$

**Limiting case 1: weak coupling**: here, we assess whether [7] recovers the correct form in the limit of weak coupling; i.e., when the amplitudes of the oscillators are constant.

Let us first consider how [8] is written when the amplitude lies on its limit cycle ($r = r_W$):

$$\frac{dr}{dt} = r_W \sum_j u_j B^j + Cu\cos\theta \quad [9]$$

We see that, despite constraining the amplitude to its limit cycle, [9] is not yet describing weak coupling, as the amplitude may still vary by virtue of a non-zero extrinsic driving factor $u$. [9] demonstrates a special case, in which there is weak intrinsic coupling within a system (via the $A$ matrix), but potentially strong coupling between the system and the external world (via the $B$ and $C$ matrices). In order to describe the weak coupling approximation, we see from [8] that if $\frac{dr}{dt}$ is zero $\forall\ \theta$, the extrinsic driving factor, $u$, must also be zero.

Setting $u = 0$ in [8] we obtain the following:

$$\frac{d\theta}{dt} = \Omega + \sum_j A^j \sin(\theta_j - \theta) \quad [10]$$

Where [10] takes the form of the Kuramoto model, differing only by the use of the DCM $A$ matrix in place of the normalized global coupling constant K/N (17).

**Limiting case 2: single-region intrinsic activity:** here, we examine equations [7] and [8] in the limit of a single region ($A = -a_{11}$) with zero external input ($u = 0$).

Such a system is described by [8] as follows:

$$\frac{dr}{dt} = -a_{11}(r - r_W) \qquad [11]$$

Solving [11]:

$$r(t) = c_1 e^{-a_{11}t} + r_W \qquad [12]$$

Where $c_1$ is a constant.

If given sufficient time, in the absence of external stimuli, activity in the isolated region will recover periodic behaviour, such that:

$$\lim_{t \to \infty} r(t) = c_2 cos(\theta) \qquad [13]$$

Under the same conditions, equation [7] reduces to:

$$\frac{d\theta}{dt} = \Omega \qquad [14]$$

Solving [14]:

$$\theta(t) = \Omega t + \emptyset \qquad [15]$$

Where $\emptyset$ is a constant. Using [12], [13] and [15], we recover a harmonic oscillator with transient and steady-state regimes.

$$r(t) = c_1 e^{-a_{11}t} + c_2 cos(\Omega t + \emptyset) \qquad [16]$$

**Simulation of coupled pendula in Unity3D**

A customisable simulation of three coupled pendula is implemented using the physics simulator as part of the Unity3D gaming engine (version 2017.3.1f1). The full implementation and underlying code are available here: https://github.com/Phase-Amplitude-Coupling/Code.git.

The simulation consists of three coupled pendula, each of which consists of a bob that is connected by a fixed joint to a rigid body cylinder, which in turn is connected by a hinged joint to a stationary hook. In addition, the cylinders are coupled to each other by spring joints. The bob of the first pendulum receives a force that is applied perpendicularly to the hinge joint and follows a Gaussian curve for the first 5 seconds of each simulation, with the same parameters used for the model-generated data (https://github.com/Phase-Amplitude-Coupling/Code.git). Each simulation is run for a total of 30 seconds at a sampling rate of 60Hz. The spring coupling strengths are initialised at a value of 0.12 for the phase-only coupling scenario and increased in steps of 0.01 to a value of 2.00 for the phase-amplitude coupling scenario.

**Bayesian model inversion and model reduction in SPM**

Fitting the models to data involves first arranging data into matrices of number of regions × time points. The sizes of these matrices are a) 3×128 for simulated phase-only and phase-amplitude coupling data, b) 3×163 for simulated coupled pendula at each coupling strength, c) 2×149 repeated across 20 non-overlapping equal-length time windows for neural data at each anaesthetic depth, and d) 5×60 for GDP data for pre- and post-WWII eras.

Given these data we estimate the optimal Bayesian states ([7] and [8]), parameters (A and C) and hyperparameters (variance of states and parameters) using spm_DEM. This inversion routine uses a Laplace approximation of states and parameters (multivariate Gaussians) in generalised coordinates of motion. That is, we estimate not only the rates of change of phase and amplitude ([7] and [8]) but also higher derivatives. We apply 4 embedding dimensions, which accommodates analytic (i.e., smooth) noise processes, unlike the martingale assumptions used in traditional Kalman filters (33). The optimisation scheme employs a dynamic expectation maximization (DEM) algorithm. The D step corresponds to state estimation using variational Bayesian filtering in generalised coordinates (that can be regarded as an instantaneous gradient descent in a moving frame of reference). The E step uses gradient ascent on the negative free energy to estimate model parameters and the M step then does the same for the hyperparameters (i.e., precision components of random fluctuations on the states and observation noise) (21). Although DEM is usually portrayed as an expectation maximisation scheme, it is in fact a variational Bayesian scheme with three steps that optimise an approximate posterior over three unknown quantities (i.e., states, parameters and hyper parameters). Specific priors are required for this inversion scheme: we used prior values of -3 for the log off-diagonal coupling parameters, with a prior variance of 1. Finally, we used a hyperprior of 6 for the log precision over observation noise.

Having applied the optimization to the model comprising both phase and amplitude states ([7] and [8]], we then use a post-hoc method (34) to estimate the evidence for the reduced models. The latter are specified by setting the prior variance over the off-diagonal parameter values of the amplitude, phase, or both amplitude and phase connectivity matrices to zero. This returns the free energy approximation to model evidence for the reduced models and the reduced posterior density over parameters. Posterior model probabilities are then derived by applying a sigmoidal softmax function $\left(\frac{1}{1+e^{\Sigma F}}\right)$ to the free energy bound on log evidence.

## Code availability

All code used in this study is available at https://github.com/Phase-Amplitude-Coupling/Code.git.

## Data availability

All data generated for the coupled pendula simulation are available at https://github.com/Phase-Amplitude-Coupling/Code.git.

**Acknowledgements**

E.D.F. and R.L were funded by the Medical Research Council (Ref: MR/R005370/1). R.J.M was funded by the Wellcome/EPSRC Centre for Medical Engineering (Ref: WT 203148/Z/16/Z). I.R.V. was funded by the Wellcome Trust (Ref: 103045/Z/13/Z). K.J.F. was funded by a Wellcome Principal Research Fellowship (Ref: 088130/Z/09/Z)

**Author contributions**

E.D.F., R.J.M., I.R.V., R.L. and K.J.F. designed and performed research, analysed data and wrote the paper.

**Competing interests**

The authors declare no competing interests.